\documentclass{article}

\usepackage{graphicx}

\usepackage{amssymb}

\begin{document}

\title{Scattering by infinitely rising one-dimensional potentials}

\author{\ \\
E.M. Ferreira, \\  \  \\
{\em Instituto de F\'{\i}sica, Universidade Federal do Rio de Janeiro,} \\
{\em 21941-972, Rio de Janeiro, Brasil}\\ \ \\
and \\ \ \\
J. Sesma\thanks{Corresponding author. Email: javier@unizar.es} \\ \ \\
{\em Departamento de F\'{\i}sica Te\'{o}rica, Facultad de Ciencias,} \\
{\em 50009, Zaragoza, Spain}}

\maketitle

\begin{abstract}
Infinitely rising one-dimensional potentials constitute impenetrable barriers which reflect totally any incident wave. However, the scattering by such kind of potentials is not structureless: resonances may occur for certain values of the energy. Here we consider the problem of scattering by the members of a family of potentials $V_a(x)=-{\rm sgn}(x)\,|x|^a$, where sgn represents the sign function and $a$ is  a positive rational number. The scattering function and the phase shifts are obtained from global solutions of the Schr\"odinger equation. For the determination of the Gamow states, associated to resonances, we exploit their close relation with the eigenvalues of the $\mathcal{PT}$-symmetric Hamiltonians with potentials $V_a^{\mathcal{PT}}(x)=-{\rm i}\,{\rm sgn}(x)\,|x|^a$. Calculation of the time delay in the scattering at real energies is used to characterize the resonances.  As an additional result, the breakdown of the $\mathcal{PT}$-symmetry of the family of potentials $V_a^{\mathcal{PT}}$ for $a<3$ may be conjectured.
\end{abstract}




\section{Introduction}

The recent progress in nano-technology has resulted in the ever increasing possibility of producing a great variety of physical systems, like quantum dots, wires, etc., many of which can be simulated by one-dimensional quantum models. Two kinds of problems are usually considered: the formation of bound states in a potential, corresponding to the trapping of a particle in a given medium, and the presence of resonances in the scattering by this potential. Here we are concerned with the second kind of problems.

In one-dimensional scattering, an incident wave is partly reflected and partly transmitted, unless the potential increases infinitely so as to constitute an impenetrable barrier, in which case the wave is totally reflected. Nevertheless, this reflection occurs after a certain dwelling, that is, with a time-delay \cite{caac} which depends on the energy of the incident wave. Peaks in the time delay may appear for certain energies, revealing the existence of resonances \cite{merz,nuss,oroz} in the scattering process.

The first indication of the presence of resonances in the scattering by an infinitely rising potential appeared in the study, by Yaris {\em et al} \cite{yari} and by Caliceti {\em et al} \cite{cali}, of the one-dimensional cubic anharmonic oscillator of Hamiltonian
\begin{equation}
H=-\,\frac{{\rm d}^2}{{\rm d}x^2}+\frac{x^2}{4}-\lambda\,x^3, \qquad \lambda>0\,,  \label{i1}
\end{equation}
and the discovery that it possesses complex eigenvalues and localized eigenfunctions. The properties of these solutions, that could be interpreted as resonances, were discussed by Alvarez \cite{alv1}. These pioneering works opened the way to further studies of resonances in anharmonic oscillators
\begin{equation}
H=-\,\frac{{\rm d}^2}{{\rm d}x^2}+\frac{x^2}{4}-g\,x^N,  \label{i2}
\end{equation}
of degree $N=3$ \cite{jen1,jen2} or of arbitrary even and odd values of $N$ \cite{zin1,zin2,jen3,prl}.

Resonances in the cubic anharmonic oscillator were intuitively associated \cite{alv1} to quantum tunneling through the potential barrier formed by the combination of the $x^2$ and $x^3$ terms in the Hamiltonian (\ref{i1}). However, this interpretation, which underlies also more recent studies of anharmonic oscillators \cite{jen1,prl}, does not seem to be satisfactory. Let us consider, for instance, the case of the Hamiltonian (\ref{i1}) with $\lambda=256$. The corresponding potential presents a barrier of maximum height $1/28311552$ between its zeros located at $x=0$ and $x=1/1024$. This imperceptible barrier cannot explain the marked localization of the wave function $\psi(x)$ of the first resonance, shown in Fig. 1. We find more plausible to associate such a localization with the cubic term in the potential. In fact, resonances have been found even in absence of the $x^2$ term in the Hamiltonian (\ref{i1}), i. e., in the case of a purely cubic potential \cite{fer2}. A reasonable explanation of the existence of such localized solutions would be the following: thinking in terms of density of probability of presence of a particle, represented by $|\psi(x)|^2$, the exponentially decreasing behavior as $x$ goes to $-\infty$ is, of course, related to the classical impenetrability of that region, whereas the decreasing probability values found for positive increasing $x$ can be associated to the fact that a classical particle in a cubic potential reaches $+\infty$ in a finite time.
\begin{figure}[t]
\resizebox{11cm}{!}{\includegraphics{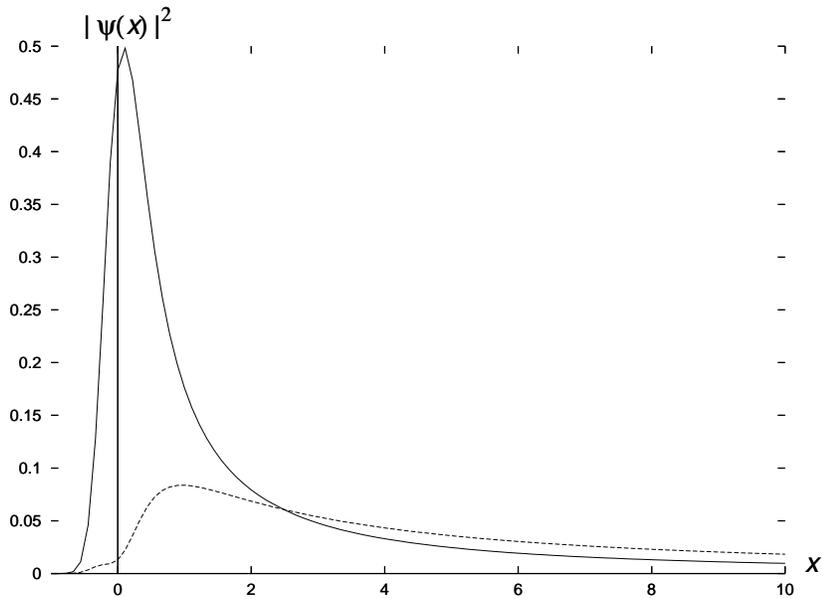}}
\caption{Squared modulus of the normalized wave functions of the first (continuous line) and second (dashed line) resonances of the Hamiltonian (\ref{i1}) with $\lambda=256$. The corresponding eigenvalues are $8.59629958291-6.24556570865$ i and $30.5501842198-22.1959889066$ i. The procedure used to compute the eigenvalues and eigenfunctions is described in  \cite{fer2}. The strong localization of the wave function of the first resonance hardly could be attributed to the existence of an extremely small barrier, of width less than $10^{-4}$ and maximum height less than $10^{-8}$, in the potential $x^2/4-256\,x^3$.} \label{cub1}
\end{figure}

Recently, resonances have been searched in one-dimensional potentials rising linearly \cite{ahm2}, quadratically \cite{ahm2,nfer}, or exponentially \cite{ahm2,ahm1,ahm3} with the distance. The algebraic solvability, in terms of well known special functions, of the Schr\"odinger equation for those potentials allows to obtain algebraic expressions for the reflection amplitude and for the time delay as a function of the energy. Resonances are clearly shown in this way.

Our main purpose in the present paper is to discuss scattering by a family of divergent-at-infinity odd potentials given by
\begin{equation}
V_a(x)=-{\rm sgn}(x)\,|x|^a \label{extra1}
\end{equation}
with positive rational values of $a$,
\begin{equation}
a=p/q\,,   \qquad p, q,\quad {\rm positive \;\; integers}.  \label{i4}
\end{equation}
An alternative notation, more convenient for considering complex scaling of the variable, is
\begin{equation}
V_a(x)=\left\{\begin{array}{lll}(-x)^a,&\qquad \mbox{for}&\quad x<0, \\ -\,x^a, &\qquad \mbox{for}&\quad x>0. \end{array}\right. \label{i3}
\end{equation}
Figure 2 illustrates the shape of these potentials for the three cases of $0<a<1$, $a=1$, and $a>1$.
Besides evaluating the phase shift experienced by a wave impinging from the right on one of those potentials, at arbitrary real energies, we are interested in the possible existence of Gamow states, that is, eigenfunctions of the Hamiltonians
\begin{equation}
H_a=-\,\frac{{\rm d}^2}{{\rm d}x^2}+V_a(x)   \label{i5}
\end{equation}
corresponding to complex eigenvalues and behaving as a pure outgoing wave for $x\to +\infty$.

Up to here, we have used the term ``resonances" to refer to such kind of solutions. This is common practice. However, in what follows, we prefer to designate them as ``Gamow states"  and to reserve the term ``resonances" for scattering states, with real energy, such that the time delay in the reflection is considerably enhanced. The occurrence of a resonance is an indication of the existence of a Gamow state of complex energy, whose real part is nearly equal to the resonant energy and whose imaginary part is small. A Gamow energy of relatively large imaginary part does not imply the existence of resonant scattering.

Very precise numerical eigenvalues of  Hamiltonians of the form $p^2/2-x^K$, with integer $K$, have been recently obtained by Fern\'andez and Garcia \cite{fega} using the complex rotation and the Riccati-Pad\'e methods. These authors, however, have not considered scattering problems.
\begin{figure}[t]
\resizebox{11cm}{!}{\includegraphics{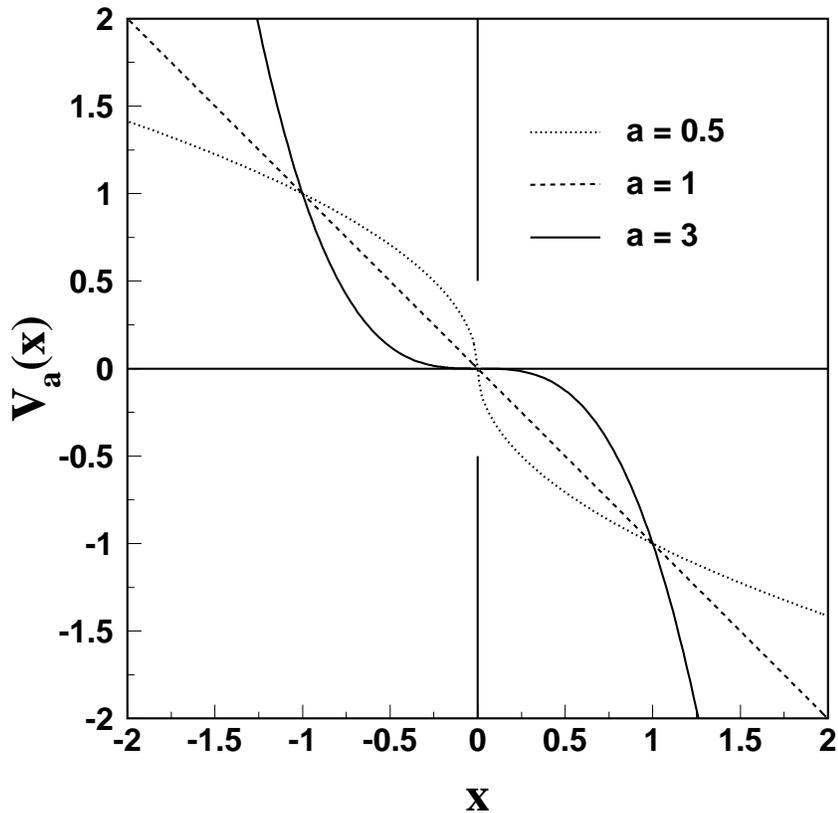}}
\caption{Three examples of potentials of the family $\{V_a(x)\}$, as given in (\ref{i3}), corresponding to each one of the three ranges of values of the parameter $0<a<1$, $a=1$, and $a>1$.} \label{potentials}
\end{figure}

Potentials of the family (\ref{i3}) present an impenetrable barrier for an incident (from the right) wave, which should be totally reflected. Therefore, the wave function representing the scattering process for one of these potentials appears, at large distances, as
\begin{equation}
\psi(x) \sim \left\{\begin{array}{ll} 0\,, &\quad x\to- \infty, \\\psi^{\rm{incoming}}(x) - S(E)\,\psi^{\rm{outgoing}}(x), & \quad x\to +\infty\,, \end{array}\right .  \label{i8}
\end{equation}
The scattering function $S(E)$, dependent on the energy $E$ of the particle associated to the incoming wave, contains all the information about the interaction of the wave with the potential. Its determination is the first purpose of this paper.

The Schr\"odinger equation to be solved is, in appropriate units for lengths and energies,
\begin{equation}
-\,\frac{{\rm d}^2\psi(x)}{{\rm d}x^2}+V_a(x)\,\psi(x)=E\,\psi(x)\,,  \qquad -\infty < x < +\infty, \label{i9}
\end{equation}
with $V_a(x)$ as given in (\ref{i3}). The two differential equations resulting for the cases of $x\leq 0$ and $x\geq 0$ can be treated simultaneously when written in the form
\begin{equation}
-\,\frac{{\rm d}^2\phi^{(\sigma)}(x)}{{\rm d}x^2}-\sigma \,x^a\,\phi^{(\sigma)}(x)=E\,\phi^{(\sigma)}(x)\,,  \quad 0\leq x < +\infty,\quad \sigma=\pm1\,, \label{i10}
\end{equation}
understanding that $\phi^{(+1)}(x)$ coincides with  $\psi(x)$ on the positive real semiaxis whereas $\phi^{(-1)}(x)$ is the mirror image, with respect to the vertical axis $x$=0, of $\psi(x)$ on the negative real semiaxis.

There are two particular values of the parameter $a$, namely $a=1$ and $a=2$, such that the corresponding Eq.~(\ref{i9}) becomes algebraically solvable. In the more general case, scattering states of the Hamiltonians  $H_a$, given
in (\ref{i5}), appear as global solutions of the Schr\"odinger equation (\ref{i9}) behaving at large distances as in Eq.~(\ref{i8}). Our method to obtain such global solutions finds inspiration in a procedure proposed by Naundorf \cite{nau1,nau2}
forty years ago. The method, later considerably improved \cite{gom1}, has been used for the solution of several quantum problems \cite{gom2,gom3,sesm}. For rational values $a>2$ a general formalism can be written. For values $0<a\leq 2$ the method is equally applicable, but the resulting expressions depend on $a$ in a manner that cannot be represented by a common form.

Section 2 recalls the algebraic solution of (\ref{i9}) for the two values of $a=1$ and $a=2$, mentioned above. Most sections of the paper are devoted to the general case of rational $a$ as given in (\ref{i4}), with the restriction $a>2$. Bases of solutions of Eq.~(\ref{i10}), useful either near the origin or at very large distances, are given in Section 3. Writing global solutions requires the knowledge of the connection factors relating the elements of the basis at the origin with those of the basis for large $x$. Section 4 explains the procedure for evaluating the connection factors. Global solutions with the appropriate behavior at infinity are then written in Section 5. This results in an expression for the scattering function and, consequently, for the phase shift in terms of the connection factors. Eigenvalues of a family of $\mathcal{PT}$-symmetric Hamiltonians $H_a^{\mathcal{PT}}$, related to $H_a$ by means of a Symanzik scaling of the variable, are reported in Section 6. The results suggest that the symmetry becomes broken for $a<3$. The mentioned Symanzik scaling, applied to the eigenvalues of $H_a^{\mathcal{PT}}$, allows to obtain immediately the Gamow energies for $H_a$.  This is done in Section 7. The possible existence of resonances associated to the Gamow energies is analyzed, also in Section 7, by computing the time delay in the scattering by $V_a$. The particular case of $a=1/2$, not included in the general case of $a>2$ and as a representative of the values of $a\in (0, 1)$, is treated in Section 8. Some final comments can be found in Section 9.

\section{Algebraically solvable cases}

For $a=1$ in our potential $V_a(x)$ given in (\ref{i3}), i. e., for the so called linear potential,
\begin{equation}
V_1(x)=-x\,, \qquad -\infty<x<+\infty\,, \label{i11}
\end{equation}
the Schr\"odinger equation (\ref{i9}), with the notation
\begin{equation}
z\equiv -(x+E)\,, \qquad u(z)\equiv \psi(x),   \label{i12}
\end{equation}
becomes the Airy equation \cite[Eq.~10.4.1]{abra}
\begin{equation}
u^{\prime\prime}(z)-z\,u(z) = 0,  \label{i13}
\end{equation}
whose solution $u(z)=$Ai$(z)$ vanishes as $z\to +\infty$ (i. e., $x\to -\infty$) and oscillates for $z\to -\infty$ (i. e., $x\to +\infty$), as expected for the physical solution. Written in terms of Hankel functions \cite[Eq.~10.4.15]{abra}, this
 physical solution, up to an arbitrary multiplicative constant, becomes
\begin{equation}
{\rm Ai}(-(x+E))=\frac{1}{2}\left(\frac{x+E}{3}\right)^{1/2}\left[{\rm e}^{{\rm i}\pi/6}H_{1/3}^{(1)}(\zeta)+{\rm e}^{-{\rm i}\pi/6}H_{1/3}^{(2)}(\zeta)\right],   \label{i14}
\end{equation}
where \begin{equation}
\zeta=\frac{2}{3}(x+E)^{3/2}\,.   \label{i15}
\end{equation}
Choosing the arbitrary multiplicative constant equal to $2\pi^{1/2}{\rm e}^{-{\rm i}\pi/4}$, and  retaining only the dominant terms in the asymptotic expansions of the Hankel functions \cite[Eqs. 9.2.7 and 9.2.8]{abra}, we obtain for the physical solution
\begin{eqnarray}
\psi_{\rm phys}(x) & = & 2\pi^{1/2}{\rm e}^{-{\rm i}\pi/4}\,{\rm Ai}(-(x+E)) \nonumber \\
&  \sim & - {\rm e}^{{\rm i}\pi/2}\,\varphi_3(x)+\varphi_4(x)\,, \qquad  {\rm for} \; x\to +\infty\,  \label{i16}
\end{eqnarray}
with
\begin{eqnarray}
\varphi_3(x) & = (x+E)^{-1/4}\exp\left({\rm i}\,\frac{2}{3}\,(x+E)^{3/2}\right)\,, \\  \varphi_4(x) & = (x+E)^{-1/4}\exp\left(-{\rm i}\,\frac{2}{3}\,(x+E)^{3/2}\right)\,.  \label{i17}
\end{eqnarray}
The Barton's criterion \cite{bart} to define incoming and outgoing waves in
divergent-at-infinity potentials shows that $\varphi_3$ and $\varphi_4$
correspond  respectively to an outgoing (to the right) and an incoming
(from the right) waves. Therefore, comparison of the asymptotic expression
of $\psi_{\rm phys}(x)$, as given in Eq.~(\ref{i16}), with that expected,
according to Eq.~(\ref{i8}), reveals that the scattering function is a constant,
\begin{equation}
S(E)= {\rm e}^{{\rm i}\pi/2}\,.   \label{i18}
\end{equation}
This lack of structure of the scattering function in the case of a one-dimensional linear potential $ V_1(x)$ has already been noticed by Ahmed {\em et al} \cite{ahm2}, who have considered a family of potentials including $V_1(x)$ as a particular case. Such lack of structure
was to be expected, since a displacement of the energy accompanied by a corresponding translation in the variable $x$ leaves invariant the Schr\"odinger equation.

The other value of $a$ leading to an algebraically solvable problem, namely $a=2$, was already considered in a previous
 paper \cite{fer1}. The solutions of Eqs. (\ref{i10}) for $a=2$
can be expressed in terms of confluent hypergeometric functions, whose asymptotic behavior is well known. The scattering function turns out to be in this case
\begin{equation}
S(E) =  \,e^{{\rm i}\pi/2}\,\frac{\mathcal{N}(E)}{\mathcal{D}(E)},   \label{i19}
\end{equation}
with the notation
\begin{eqnarray}
\mathcal{N}(E) & = & \frac{{\rm e}^{{\rm i}\pi/8}}{\Gamma\left(\frac{3-E}{4}\right)\,\Gamma\left(\frac{1+{\rm i}E}{4}\right)}
+ \frac{{\rm e}^{-{\rm i}\pi/8}}{\Gamma\left(\frac{1-E}{4}\right)\,\Gamma\left(\frac{3+{\rm i}E}{4}\right)},  \label{i20} \\
\mathcal{D}(E) & = & \frac{{\rm e}^{-{\rm i}\pi/8}}{\Gamma\left(\frac{3-E}{4}\right)\,\Gamma\left(\frac{1-{\rm i}E}{4}\right)}
+ \frac{{\rm e}^{{\rm i}\pi/8}}{\Gamma\left(\frac{1-E}{4}\right)\,\Gamma\left(\frac{3-{\rm i}E}{4}\right)}.  \label{i21}
\end{eqnarray}
A  graphical representation of this scattering function and a thorough discussion of its properties was given in \cite{fer1}.

\section{General case of $a>2$. Local solutions of the Schr\"odinger equation}

Now we consider the general case of $a$ as given in (\ref{i4}). The restriction to $a>2$ is not essential for the application of our method of solution of the Schr\"odinger equation, but it allows us to write the equations below in a common form. For $a\leq 2$, certain expressions, like the Thom\'e solutions to be defined below, cannot be written in a form independent of $a$. This is illustrated in Section 8, where the case of $a=1/2$ is considered.

To avoid dealing with fractional powers, we introduce the  change of variables
\begin{equation}
x\equiv t^{2q},\qquad \phi^{(\sigma)}(x)\equiv t^{q-1/2}\,w^{(\sigma)}(t), \qquad t\in [0,+\infty). \label{ii1}
\end{equation}
The differential equation (\ref{i10}) turns, in this way, into
\begin{equation}
t^2\,\frac{{\rm d}^2w^{(\sigma)}(t)}{{\rm d}t^2}+\left(4q^2\,\sigma\,t^{2p+4q}+4q^2\,E\,t^{4q}-q^2+1/4\right)w^{(\sigma)}(t)=0 ~.  \label{ii2}
\end{equation}
Two kinds of solutions of this equation are to be considered. The first kind is a pair of (Frobenius) solutions written as series expansions, convergent for finite $t$,
\begin{equation}
w_i^{(\sigma)}(t)=t^{\nu_i}\,\sum_{n=0}^\infty c_{n,i}^{(\sigma)}\,t^n,  \qquad i=1, 2,  \label{ii3}
\end{equation}
with indices
\begin{equation}
\nu_1= -q+1/2, \qquad \nu_2=q+1/2,  \label{ii4}
\end{equation}
and coefficients given by the recurrence relation
\[
c_{0,i}^{(\sigma)}=1, \qquad c_{2q,1}^{(\sigma)}=0,
\]
\begin{equation}
n(n+2\nu_i-1)c_{n,i}^{(\sigma)}=-\,4q^2E\,c_{n-4q,i}^{(\sigma)}-4q^2\sigma\,c_{n-2p-4q,i}^{(\sigma)} \,.  \label{ii5}
\end{equation}
The corresponding solutions
\begin{equation}
\phi_i^{(\sigma)}(x)\equiv t^{q-1/2}\,w_i^{(\sigma)}(t)\,,  \qquad i=1, 2,   \label{ii6}
\end{equation}
of Eq.~(\ref{i10}) are regular for finite $x$ and constitute an adequate basis to express the physical solution of the Schr\"odinger equation as linear combination
\begin{equation}
\psi_{\rm phys}(x)=\left\{\begin{array}{lll}A_1^{(-1)}\,\phi_1^{(-1)}(-x)+A_2^{(-1)}\,\phi_2^{(-1)}(-x),&\qquad \mbox{for}&\quad x<0, \\
 A_1^{(+1)}\,\phi_1^{(+1)}(x)+A_2^{(+1)}\,\phi_2^{(+1)}(x), &\qquad \mbox{for}&\quad x>0.\end{array}\right . \label{ii7}
\end{equation}
 The continuity at $x=0$ of $\psi_{\rm phys}(x)$ and of its derivative is guaranteed if we take
\begin{equation}
A_1^{(-1)}=A_1^{(+1)}=A_1\,, \qquad A_2^{(-1)}=-A_2^{(+1)}=-A_2\,, \label{ii8}
\end{equation}
with constants $A_1$ and $A_2$ to be determined as explained below.

The second kind of solutions of the differential equation (\ref{ii2}) to be considered is the pair of formal (Thom\'e) solutions
\begin{equation}
w_j^{(\sigma)}(t)=\exp \left[\alpha_j^{(\sigma)}\,t^{p+2q}/(p+2q)\right]\,t^{\mu_j}\,\sum_{m=0}^\infty a_{m,j}^{(\sigma)}\,t^{-m}, \qquad j=3, 4, \label{ii9}
\end{equation}
with exponents
\begin{equation}
\alpha_3^{(\sigma)}=-\alpha_4^{(\sigma)}=-2q(-\sigma)^{1/2}\,, \qquad \mu_3=\mu_4=-(p+2q-1)/2, \label{ii10}
\end{equation}
and coefficients given by
\begin{eqnarray}
a_{0,j}^{(\sigma)} =  1, \qquad
2\alpha_j^{(\sigma)}ma_{m,j}^{(\sigma)} & = & 4q^2E\,a_{m-p+2q,j}^{(\sigma)} \nonumber  \\
 & &\hspace{-30pt} +\,\left((m\! +\!\mu\! -\! 1)(m\! +\! \mu)-q^2+1/4\right)a_{m-p-2q,j}^{(\sigma)}\,.  \label{ii11}
\end{eqnarray}
The series of negative powers of $t$ in the right-hand side of Eq.~(\ref{ii9}) is
 non-convergent in general, but it becomes asymptotic for $t\to+\infty$. The symbol $(-\sigma)^{1/2}$ entering the definition (\ref{ii10}) of the exponents $\alpha_j^{(\sigma)}$ is two-valued. To avoid ambiguities, we adopt the convention
\begin{equation}
(-\sigma)^{1/2} = \left\{\begin{array}{lll} 1  \qquad {\rm when} \quad \sigma=-1\,,  \\
 -{\rm i}  \qquad {\rm when} \quad \sigma=+1\,. \end{array}.\right.  \label{ii12}
\end{equation}
Then, with the notation
\begin{equation}
\phi_j^{(\sigma)}(x)\equiv t^{q-1/2}\,w_j^{(\sigma)}(t), \qquad \sigma=\pm 1, \quad j=3, 4, \label{ii13}
\end{equation}
it is evident that, for real energy $E$, the wave function $\psi(x)$ corresponding to $\phi_3^{(-1)}$ or $\phi_4^{(-1)}$ respectively vanishes or increases exponentialy as $x\to -\infty$. On the other hand, bearing in mind that the flux of probability associated to a wave function $\psi(x)$ is given (in appropriate units) by
\begin{equation}
j(x) = -\,{\rm i}\left(\psi(x)^*\,\frac{{\rm d}\psi(x)}{{\rm d}x}-\frac{{\rm d}\psi(x)^*}{{\rm d}x}\,\psi(x)\right),  \label{ii14}
\end{equation}
where the asterisk stands for complex conjugation, one realizes immediately that $\phi_3^{(+1)}(x)$ and $\phi_4^{(+1)}(x)$, which for real $E$ are complex conjugate of each other, correspond respectively to outgoing (to the right) and incoming (from the right) waves.

\section{Connection factors}

Written the physical solution of the Schr\"odinger equation in the form (\ref{ii7}), the constants $A_1$ and $A_2$ are to be chosen so as to ensure that $\psi_{\rm phys}$ is regular at $x\to\pm\infty$.
The behavior of the Frobenius solutions, $\phi_i^{(\sigma)}$ ($i=1, 2$), as $x\to\pm\infty$ can be written if their connection factors $T_{i,j}^{(\sigma)}$ with the Thom\'e solutions, defined by the relations
\begin{equation}
\phi_i^{(\sigma)}(x)\sim T_{i,3}^{(\sigma)}\,\phi_3^{(\sigma)}(x)+T_{i,4}^{(\sigma)}\,\phi_4^{(\sigma)}(x)\,, \qquad x\to \infty, \qquad i=1, 2\,,  \label{iii1}
\end{equation}
are known. The computation of the connection factors, for each given value of the energy $E$, is the crucial point in our method of solution of the Schr\"odinger equation. We use a procedure, whose theoretical basis was given in \cite{gom1}.

Obviously, the connection factors can be written as the quotient of two Wronskians
\begin{equation}
 T_{i,3}^{(\sigma)}=\frac{\mathcal{W}\left[ \phi_i^{(\sigma)},\,\phi_4^{(\sigma)}\right]}{\mathcal{W}\left[ \phi_3^{(\sigma)},\,\phi_4^{(\sigma)}\right]},\qquad
 T_{i,4}^{(\sigma)}=\frac{\mathcal{W}\left[ \phi_i^{(\sigma)},\,\phi_3^{(\sigma)}\right]}{\mathcal{W}\left[ \phi_4^{(\sigma)},\,\phi_3^{(\sigma)}\right]},\qquad  i=1,2, \label{iii2}
\end{equation}
of the solutions discussed above. The denominators, easily obtained by direct evaluation, are
\begin{equation}
\mathcal{W}\left[\phi_3^{(\sigma)},\,\phi_4^{(\sigma)}\right]=-\mathcal{W}\left[\phi_4^{(\sigma)},\,\phi_3^{(\sigma)}\right]=\alpha_4^{(\sigma)}/q\,.
\label{iii3}
\end{equation}
The computation of the numerators is more involved. In principle,
\begin{equation}
\mathcal{W}\left[ \phi_i^{(\sigma)},\,\phi_j^{(\sigma)}\right]=\frac{1}{2q}\,\mathcal{W}\left[ w_i^{(\sigma)},\,w_j^{(\sigma)}\right]\,,
\label{iii4}
\end{equation}
the right-hand side being independent of $t$. However, their direct evaluation, with the expressions (\ref{ii3}) and (\ref{ii9}) for $w_i$ and $w_j$, produces a doubly infinite series of positive and negative powers of $t$, from which it is not clear how to obtain its constant value. Our procedure to overcome this difficulty consists of comparing the asymptotic expansions of two conveniently chosen functions, $f(t)$ and $h(t)$, such that
\begin{equation}
f(t)=\mathcal{W}\left[w_i^{(\sigma)},\,w_j^{(\sigma)}\right]\,h(t)\,.  \label{iii5}
\end{equation}
The uniqueness of the asymptotic expansion of any function \cite[Sec. 1.3]{erde}, for a given asymptotic sequence, allows one to obtain the value of $\mathcal{W}[w_i^{(\sigma)},\,w_j^{(\sigma)}]$. For the function $f(t)$ we choose the Wronskian of two auxiliary functions
\begin{eqnarray}
u_{i,j}^{(\sigma)}(t)=&\exp\left( -\alpha_j^{(\sigma)}\,t^{p+2q}/(2p+4q)\right)\,w_{i}^{(\sigma)}(t),  \nonumber \\
u_j^{(\sigma)}(t)=&\exp\left( -\alpha_j^{(\sigma)}\,t^{p+2q}/(2p+4q)\right)\,w_j^{(\sigma)}(t),   \nonumber \\
  &  i=1, 2,  \qquad  j=3, 4 .  \label{iii6}
\end{eqnarray}
The corresponding function $h(t)$ becomes evident from the obvious relation
\begin{equation}
\mathcal{W}\left[u_{i,j}^{(\sigma)},u_j^{(\sigma)}\right]=\mathcal{W}\left[w_{i}^{(\sigma)},w_j^{(\sigma)}\right] \,\exp\left(-\,\alpha_j^{(\sigma)}\,t^{p+2q}/(p+2q)\right). \label{iii7}
\end{equation}
A formal expansion of the left-hand side of this equation can be obtained by direct computation, bearing in mind the definitions (\ref{iii6}) and the expressions (\ref{ii3}) and (\ref{ii9}). The result can be written in the form
\begin{equation}
\mathcal{W}\left[u_{i,j}^{(\sigma)},u_j^{(\sigma)}\right] \sim \sum_{n=-\infty}^\infty \eta_{n,i,j}^{(\sigma)}\,t^{n+\nu_i+\mu_j},   \label{iii8}
\end{equation}
with the notation
\begin{equation}
  \eta_{n,i,j}^{(\sigma)} = \sum_{m=0}^\infty a_{m,j}^{(\sigma)}\Big(\alpha_j^{(\sigma)}\,c_{n+m+1-p-2q,i}^{(\sigma)}
-(n\!+\!2m\!+\!1\!+\!\nu_i\!-\!\mu_j)\,c_{n+m+1,i}^{(\sigma)}\Big).    \label{iii9}
\end{equation}
In order to write a similar formal expansion for the right-hand side of the relation (\ref{iii7}), we write the Wronskian of $w_{i}^{(\sigma)}$ and $w_j^{(\sigma)}$ as the sum of $p+2q$ unknown constants, $\{g_{L,i,j}^{(\sigma)}\}$ ($L=0, 1, \ldots ,p+2q-1$),
\begin{equation}
\mathcal{W}\left[w_i^{(\sigma)},w_j^{(\sigma)}\right]=\sum_{L=0}^{p+2q-1} g_{L,i,j}^{(\sigma)}\,.  \label{iii10}
\end{equation}
This allows to write Eq.~(\ref{iii7}) in the form
\begin{equation}
\sum_{n=-\infty}^\infty \eta_{n,i,j}^{(\sigma)}\,t^{n+\nu_i+\mu_j}\sim\sum_{L=0}^{p+2q-1} \left(g_{L,i,j}^{(\sigma)}\,\exp\left(-\alpha_j^{(\sigma)}\,t^{p+2q}/(p+2q)\right)\right)\,.
\label{iii11}
\end{equation}
Now, recalling the Heaviside's exponential series \cite[Sec. 2.12]{hard},
\begin{equation}
\exp(z)\sim\sum_{n=-\infty}^{\infty}\frac{z^{n+\delta}}{\Gamma(n+1+\delta)},  \qquad  |\arg(z)|<\pi\,,   \label{iii12}
\end{equation}
valid for arbitrary $\delta$, we can write adequate formal expansions
\begin{equation}
\exp\left(-\alpha_j^{(\sigma)}\,t^{p+2q}/(p+2q)\right)\sim\sum_{n=-\infty}^{\infty}
\frac{\left(-\alpha_j^{(\sigma)}\,t^{p+2q}/(p+2q)\right)^{n+\delta_{L,i,j}}}{\Gamma(n+1+\delta_{L,i,j})},    \label{iii13}
\end{equation}
with
\begin{equation}
\delta_{L,i,j}=(\nu_i+\mu_j+L)/(p+2q), \qquad L=0, 1, \ldots, p+2q-1, \label{iii14}
\end{equation}
for the exponentials appearing in each one of the $p+2q$ terms in the right-hand side of Eq.~(\ref{iii11}), which becomes
\begin{equation}
\sum_{m=-\infty}^\infty \eta_{m,i,j}^{(\sigma)}t^{m+\nu_i+\mu_j} \sim \sum_{L=0}^{p+2q-1} g_{L,i,j}^{(\sigma)}\sum_{n=-\infty}^{\infty}\frac{\left(-\alpha_j^{(\sigma)}t^{p+2q}/(p+2q)\right)^{n+\delta_{L,i,j}}}
{\Gamma(n+1+\delta_{L,i,j})}.  \label{iii15}
\end{equation}
Comparison of the coefficients of $t^{(p+2q)n+\nu_i+\mu_j+L}$ ($L=0,1,\ldots,p+2q-1$) in
the two sides of this equation allows one to determine the unknown constants $g_{L,i,j}^{(\sigma)}$ and, according to (\ref{iii10}), to obtain
\begin{equation}
\mathcal{W}\left[w_{i}^{(\sigma)},w_j^{(\sigma)}\right]=\sum_{L=0}^{p+2q-1}\frac{\Gamma(n+1+\delta_{L,i,j})}
{\left(-\alpha_j^{(\sigma)}/(p+2q)\right)^{n+\delta_{L,i,j}}}\; \eta_{(p+2q)n+L,i,j}^{(\sigma)}\,.    \label{iii16}
\end{equation}
The condition $|\arg(z)|<\pi$ for the validity of the representation (\ref{iii12}) entails,  for the validity of the expansions (\ref{iii13}), the condition $|\arg(-\alpha_j^{(\sigma)})|<\pi$. This can be fulfilled, with an adequate representation of the minus sign in front of $\alpha$ by ${\rm e}^{{\rm i}\pi}$ or ${\rm e}^{-{\rm i}\pi}$, for all values of the indices $\sigma$ and $j$, except for $\sigma=-1$ and $j=4$, since $\arg\left(\alpha_4^{(-1)}\right)=0$. In this case, the positive real semi-axis is a Stokes ray for $T_{i,3}^{(-1)}$ and, as usual, this connection factor should be obtained as the average of its values in the adjacent Stokes sectors. The same effect is obtained if we take for the Wronskian of $w_{i}^{(-1)}$ and $w_4^{(-1)}$ on the ray $\arg t=0$ the average of its expressions for $t$ just above and below the positive real semiaxis. We obtain, in this way,
\begin{eqnarray}
\mathcal{W}\left[w_{i}^{(-1)},w_4^{(-1)}\right] & = & (-1)^n\sum_{L=0}^{p+2q-1}\Bigg[\frac{\cos (\delta_{L,i,4}\pi)\,\Gamma(n+1+\delta_{L,i,4})}
{\left(\alpha_4^{(-1)}/(p+2q)\right)^{n+\delta_{L,i,4}}} \nonumber \\
& & \hspace{100pt}\times\,\eta_{(p+2q)n+L,i,4}^{(-1)}\Bigg]. \label{iii17}
\end{eqnarray}

Once the necessary  Wronskians have been calculated, it is immediate to obtain, from Eq.~(\ref{iii2}), the connection factors.
The values of the $\mathcal{W}[w_{i}^{(\sigma)},w_j^{(\sigma)}]$ are independent of the integer $n$ appearing in the right-hand sides of their expressions (\ref{iii16}) and (\ref{iii17}). Different choices for $n$ must lead to the same values of the Wronskians. Another
useful test of the accuracy of the computed values of the connection factors is the relation
\begin{equation}
T_{1,3}^{(\sigma)}\,T_{2,4}^{(\sigma)}-T_{2,3}^{(\sigma)}\,T_{1,4}^{(\sigma)} =
\frac{\mathcal{W}\left[\phi_1^{(\sigma)},\phi_2^{(\sigma)}\right]}{\mathcal{W}\left[\phi_3^{(\sigma)},\phi_4^{(\sigma)}\right]} =
q \,(\alpha_4^{(\sigma)})^{-1},   \label{iii18}
\end{equation}
which must be satisfied.

\section{Global solution of the Schr\"odinger equation}

Written the physical solution in the form (\ref{ii7}), its behavior in the far regions is given, in view of the relations (\ref{ii8}) and (\ref{iii1}), by
\begin{equation}
  \psi_{\rm phys}(x)\sim\left\{\begin{array}{ll}\left(A_1T_{1,3}^{(-1)}-A_2T_{2,3}^{(-1)}\right)\,\phi_3^{(-1)}(-x) \\ \hspace{20pt} +\left(A_1T_{1,4}^{(-1)}-A_2T_{2,4}^{(-1)}\right)\,\phi_4^{(-1)}(-x),& \mbox{as}\quad x\to -\infty, \\
\left(A_1T_{1,3}^{(+1)}+A_2T_{2,3}^{(+1)}\right)\,\phi_3^{(+1)}(x) \\  \hspace{20pt} +\left(A_1T_{1,4}^{(+1)}+A_2T_{2,4}^{(+1)}\right)\,\phi_4^{(+1)}(x), &\mbox{as}\quad x\to +\infty,\end{array}\right. \label{iv1}
\end{equation}
The necessary cancelation of the divergent term as $x\to -\infty$ imposes on the coefficients $A_1$ and $A_2$ the restriction
\begin{equation}
A_1\,T_{1,4}^{(-1)}-A_2\,T_{2,4}^{(-1)}=0\,.  \label{iv2}
\end{equation}
This condition fixes the coefficients $A_1$ and $A_2$ up to a common arbitrary multiplicative constant. We may, therefore, add the requirement
\begin{equation}
A_1\,T_{1,4}^{(+1)}+A_2\,T_{2,4}^{(+1)}=1\,,  \label{iv3}
\end{equation}
unless the connection factors be such that
\begin{equation}
T_{1,4}^{(-1)}\,T_{2,4}^{(+1)}+T_{2,4}^{(-1)}\,T_{1,4}^{(+1)}=0\,,  \label{iv4}
\end{equation}
in which case
\begin{equation}
A_1\,T_{1,4}^{(+1)}+A_2\,T_{2,4}^{(+1)}=0  \label{iv5}
\end{equation}
and $\psi_{\rm phys}(x)$ becomes a pure outgoing wave as $x\to\infty$.
Leaving aside, for the moment, this possibility, which corresponds to a Gamow state, we obtain for the coefficients $A_1$ and $A_2$ the expressions
\begin{equation}
A_1=\frac{T_{2,4}^{(-1)}}{T_{1,4}^{(-1)}\,T_{2,4}^{(+1)}+T_{2,4}^{(-1)}\,T_{1,4}^{(+1)}}\,, \quad     A_2=\frac{T_{1,4}^{(-1)}}{T_{1,4}^{(-1)}\,T_{2,4}^{(+1)}+T_{2,4}^{(-1)}\,T_{1,4}^{(+1)}}\,. \label{iv6}
\end{equation}
Comparison of the expression (\ref{iv1}) with the expected behavior, Eq.~(\ref{i8}), of the physical solution gives for the scattering function
\begin{equation}
S(E)=-\, \frac{N(E)}{D(E)}\,,  \label{iv7}
\end{equation}
where
\begin{equation}
\begin{array}{ll}N(E) = T_{1,4}^{(-1)}\,T_{2,3}^{(+1)}+T_{2,4}^{(-1)}\,T_{1,3}^{(+1)}\,, \\
D(E) = T_{1,4}^{(-1)}\,T_{2,4}^{(+1)}+T_{2,4}^{(-1)}\,T_{1,4}^{(+1)}\,. \end{array}   \label{iv8}
\end{equation}
From the structure of the connection factors one can see that
\begin{equation}
N(E^*)=[D(E)]^*\,, \qquad {\rm and} \qquad  D(E^*)=[N(E)]^*\,,   \label{iv9}
\end{equation}
which imply the fulfilment of the familiar unitarity condition
\begin{equation}
S(E)[S(E^*)]^*=1\,.   \label{iv10}
\end{equation}
This expression becomes $|S(E)|=1$ when the energy $E$ is real. In this case, it is possible to define a (real) phase shift, $\delta(E)$, such that
\begin{equation}
S(E)=\exp[2{\rm i}\,\delta(E)]\,.   \label{iv11}
\end{equation}

We show in Fig. 3 the phase shift for energies in the interval $-5<E<15$ for several values of $a$. When comparing the steepness of the different curves, one should bear in mind that the variable $E$ represented on the horizontal axis is not the physical energy, but a dimensionless parameter proportional to it. If the physical problem is that of a particle of mass $m$ and a potential $V_0x^a$, the proportionality constant is, obviously,  $(2m/\hbar^2)^{a/(a+2)}V_0^{-2/(a+2)}$.
\begin{figure}
\resizebox{11cm}{!}{\includegraphics{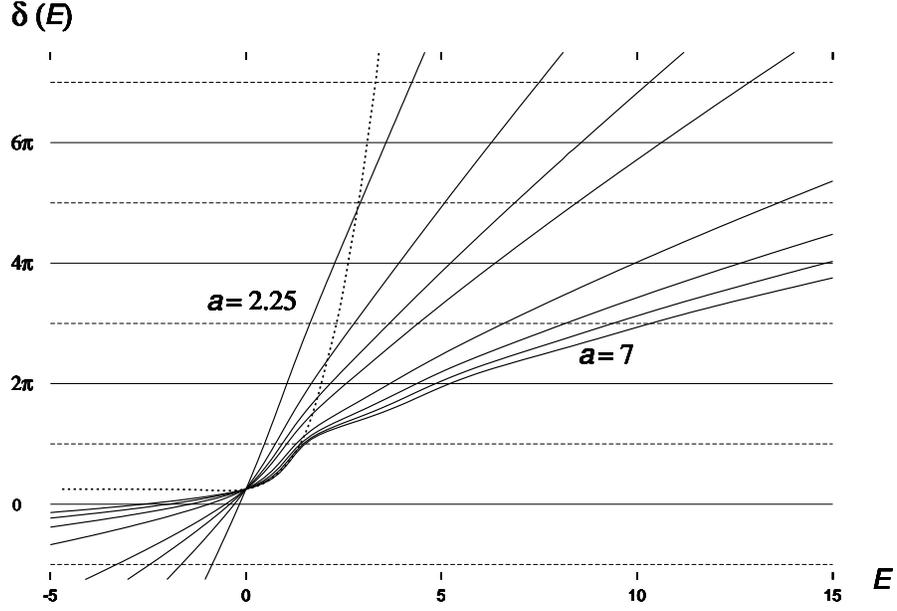}}
\caption{Phase shift of a wave scattered by the potential (\ref{i3}) with some values of $a$, namely $a=$2.25, 2.5, 2.75, 3, 4, 5, 6 and 7 (continuous lines), and a=0.5 (dotted line).}
\label{phsha=v}
\end{figure}

\section{Related $\mathcal{PT}$-symmetric potentials}

Now we turn our attention to the Gamow states. Their energies are the (complex) values of $E$ for which Eq.~(\ref{iv4}) is fulfilled. Alternatively, we may benefit from the close connection between the eigenstates of Hamiltonians (\ref{i5}) and those of the $\mathcal{PT}$-symmetric ones
\begin{equation}
H_a^{\mathcal{PT}}=-\,\frac{{\rm d}^2}{{\rm d}x^2}+V_a^{\mathcal{PT}}(x)\,,   \label{i6}
\end{equation}
with
\begin{equation}
V_a^{\mathcal{PT}}(x)=-{\rm i}\,{\rm sgn}(x)\,|x|^a\,,
\end{equation}
or, equivalently,
\begin{equation}
V_a^{\mathcal{PT}}(x)=\left\{\begin{array}{lll}{\rm i}\,(-x)^a,&\qquad \mbox{for}&\quad x<0, \\ -\,{\rm i}\,x^a, &\qquad \mbox{for}&\quad x>0,\end{array}\right.\,. \label{i7}
\end{equation}
Such connection is well known from the birth of the $\mathcal{PT}$-symmetric quantum theory marked by the seminal paper of Bender and Boettcher \cite{ben1}. (A tutorial introduction to $\mathcal{PT}$-symmetry and a later complete update can be found in two other papers by Bender \cite{ben2,ben3}.) In fact, the change of variable (Symanzik scaling)
\begin{equation}
x=\exp[{\rm i}\pi/(2a+4)]\,\hat{x}, \label{v1}
\end{equation}
converts the Schr\"odinger equation (\ref{i9}) in an analogous one, for the variable $\hat{x}$, with the potential
\begin{equation}
V_a^{\mathcal{PT}}(\hat{x})=\left\{\begin{array}{lll}{\rm i}\,(-\hat{x})^a,&\qquad \mbox{for}&\quad\hat{x}<0, \\ -{\rm i}\,\hat{x}^a, &\qquad \mbox{for}&\quad \hat{x}>0,\end{array}\right. \label{v2}
\end{equation}
and energy
\begin{equation}
E^{\mathcal{PT}}= \exp({\rm i}\pi/(a+2))\,E\,.  \label{v3}
\end{equation}
Besides, under such complex scaling, the solution corresponding to a Gamow state (vanishing incoming wave) of the Hamiltonian (\ref{i5}) becomes a bound state (vanishing wave function at infinity) in the potential (\ref{v2}). In other words, the positions in the lower complex $E$ half-plane of the poles of the scattering function (Gamow states) for the potential (\ref{i1}) are reached by a rotation of angle $-\pi/(a+2)$ of the poles on the real $E^{\mathcal{PT}}$ axis (bound states) of the scattering function for the potential (\ref{i7}).

We have, therefore, two strategies at our disposal. The first one is to find the complex $E$ zeros of the left-hand side of Eq.~(\ref{iv4}). The second one requires the solution of
the Schr\"odinger equation for the Hamiltonian (\ref{i6}) (analogous to Eq.~(\ref{i10}), with $\sigma$ replaced by i$\sigma$ and $E$ by $E^{\mathcal{PT}}$), to compute their connection factors $T_{i,j}^{\mathcal{PT}(\sigma)}$ by using the procedure described in Sections 3 and 4, to find the zeros $E^{\mathcal{PT}}$ of
\[
T_{1,4}^{\mathcal{PT}(-1)}\,T_{2,4}^{\mathcal{PT}(+1)}+T_{2,4}^{\mathcal{PT}(-1)}\,T_{1,4}^{\mathcal{PT}(+1)}=0\,,
\]
and, finally, to apply the inverse scaling
\begin{equation}
E= \exp(-{\rm i}\pi/(a+2))\,E^{\mathcal{PT}}.  \label{v4}
\end{equation}
to obtain the Gamow energies. In both strategies one has to find zeros of a function computed numerically. But it is much easier to determine real zeros (in the case of unbroken $\mathcal{PT}$-symmetry) than complex ones. For this reason we have adopted the second strategy. As a bonus, we unveil the breakdown of the $\mathcal{PT}$-symmetry, that is, the existence of complex eigenvalues, in the family of Hamiltonians $H_a^\mathcal{PT}$ for $a<3$.

We report in Table 1 the five lowest eigenvalues of the $PT$-symmetric Hamiltonians (\ref{i6}), for some values of $a$. For comparison, we have added the eigenvalues of $H_2^{\mathcal{PT}}$, for which the Schr\"odinger equation becomes algebraically solvable \cite{fer3}. The three lowest eigenvalues of $H_5^{\mathcal{PT}}$ and $H_7^{\mathcal{PT}}$, obtained by using the Riccati-Pad\'{e} method, are shown in Table 6 of \cite{fega}. The total agreement with our results makes evident the power of the method.
\begin{table}
\caption {The first five eigenvalues $E^{\mathcal{PT}}$ of the Hamiltonians (\ref{i6}) for several values of the parameter $a$. The $\mathcal{PT}$-symmetry of $H_a$ for $a\geq 3$ is unbroken, as its spectrum is entirely real. For the considered values of $a<3$ the number of real eigenvalues is finite and pairs of complex eigenvalues appear: the $\mathcal{PT}$-symmetry becomes broken. (In the case of $a=2.75$, the first complex conjugate pair is constituted by the 6th and 7th eigenvalues, as explained in the text.) The mechanism of convergence of real eigenvalues into complex ones, as $a$ decreases from 3, is that shown in \cite{ben1} and \cite{ben2}.}
\begin{tabular}{|l|rrrrr|}
\hline
$a$ & 1st eigenv. & 2nd eigenv. & 3rd eigenv. & 4th eigenv. & 5th eigenv.  \\
\hline
\footnotesize 2 & \footnotesize 1.2580917622 & \multicolumn{2}{c}{\footnotesize 4.99131440 $\pm$ 0.78048559 i} & \multicolumn{2}{c|}{\footnotesize 8.61814319 $\pm$ 3.36325532 i} \\
\footnotesize 2.125 & \footnotesize 1.2339159563 & \footnotesize 4.8245841893 & \footnotesize 5.5401494915 & \multicolumn{2}{c|}{\footnotesize 9.30471749 $\pm$ 3.04996251 i} \\
\footnotesize 2.2 & \footnotesize 1.2217833975 & \footnotesize 4.5449222689 & \footnotesize 6.0175212001 & \multicolumn{2}{c|}{\footnotesize 9.71611618 $\pm$ 2.82567175 i} \\
\footnotesize 2.25 & \footnotesize 1.2145298920 & \footnotesize 4.4495432370 & \footnotesize 6.2362764414 & \multicolumn{2}{c|} {\footnotesize 9.98930212 $\pm$ 2.65983187 i} \\
\footnotesize 2.5 & \footnotesize 1.1862679202 & \footnotesize 4.2168282738 & \footnotesize 6.9332328859 & \multicolumn{2}{c|} {\footnotesize 11.31541913 $\pm$ 1.56603260 i$\;$} \\
\footnotesize 2.75 & \footnotesize 1.1678900359 & \footnotesize 4.1350219842 & \footnotesize 7.3113845045 &\footnotesize 11.2251446520 & \footnotesize 13.7120060272 \\
\footnotesize 3 & \footnotesize 1.1562670720 & \footnotesize 4.1092287528 & \footnotesize 7.5622738550 & \footnotesize 11.3144218202 & \footnotesize 15.2915537504 \\
\footnotesize 4 & \footnotesize 1.1468501770 & \footnotesize 4.1943612352 & \footnotesize 8.3020614579 & \footnotesize 12.9101152346 & \footnotesize 18.1061761121 \\
\footnotesize 5 & \footnotesize 1.1647704080 & \footnotesize 4.3637843677 & \footnotesize 8.9551669982 & \footnotesize 14.4177548303 & \footnotesize 20.6101375101 \\
\footnotesize 6 & \footnotesize 1.1928491725 & \footnotesize 4.5449872159 & \footnotesize 9.5456178719 & \footnotesize 15.7278538420 & \footnotesize 22.8656231475 \\
\footnotesize 7 & \footnotesize 1.2247116893 & \footnotesize 4.7214625354 & \footnotesize 10.0754495631 & \footnotesize 16.8724570744 & \footnotesize 24.8575115670 \\
\hline
\end{tabular}
\end{table}

Table 1 also illustrates the breakdown of $\mathcal{PT}$-symmetry for $a$ below a critical value, $a_{\rm cr}$, which we conjecture to be equal to 3. For values, $a\geq 3$, the spectrum of $H_a$ is entirely real, according to  \cite{ben1} and \cite{shin}. In the case of $a=3$, for instance, besides the first five eigenvalues shown in table 1, we have obtained the sequence
\[
\begin{array}{ll}19.4515291307\,, \quad 23.7667404355\,, \quad 28.2175249730\,, \quad  32.7890827819\,, \\
                 37.4698253605\,, \quad 42.2504052192\,, \quad 47.1231055739\,,  \quad 52.0814360527\,, \quad {\rm etc.} \end{array}
\]
For $a=2.75$, instead, only the first five eigenvalues, shown in Table 1, are real; then
there is the  complex conjugate pair 18.91286864 $\pm$ 2.84850650 i\,. For $a=2.5$, only three real eigenvalues remain. And so on. The symmetry-breakdown mechanism is entirely analogous to that reported by Bender and Boettcher \cite{ben1} for the family of Hamiltonians
\begin{equation}
H_N^{\rm BB} = -\frac{{\rm d}^2}{{\rm d}x^2}-({\rm i}\,x)^N\,, \label{v5}
\end{equation}
whose $\mathcal{PT}$-symmetry becomes broken for $N<2$. (Notice that our family $\{H_a^\mathcal{PT}\}$ of Hamiltonians is different from the $\{H_N^{\rm BB}\}$ of Bender and Boettcher. However, $V_3^{\mathcal{PT}}(x)=\mathcal{P}\,V_3^{BB}(x)$ and the spectra of $H_3^{\mathcal{PT}}$ and $H_3^{BB}$ are the same. For other values of $a$ and $N$, the potentials $V_a^\mathcal{PT}$ and $V_N^{\rm BB}$ may be the same (for $a=N=5, 9, \ldots$) or $\mathcal{P}$-transformed of each other (for $a=N=7, 11, \ldots$), but the spectra of $H_a^\mathcal{PT}$ and $H_N^{\rm BB}$, with odd integer $a=N>3$, should not be expected to be coincident, since the self-adjoint extensions considered in the two cases are different: we are imposing to the eigenfunctions of $H_a^\mathcal{PT}$ boundary conditions on the real axis, whereas, in the case of $H_N^{\rm BB}$ with $N\geq 4$, the real axis is replaced by a contour in the complex plane.)

The fact that complex $\mathcal{PT}$-symmetric Hamiltonians possess real eigenvalues and, consequently, non-decaying eigenstates has been attributed \cite{nobl} to the exact balance of the positive and negative imaginary parts of the potential, that is, to the equilibrated source and drain of probability associated to those imaginary parts. But the reason of the breakdown of the  $\mathcal{PT}$-symmetry is far from being completely understood. It has been conjectured \cite{ben4} that, except in very exceptional cases, the analyticity of the potential is a necessary condition for the entire spectrum of a complex $\mathcal{PT}$-symmetric Hamiltonian to be real. In a recent paper \cite{ahm4}, Z.~Ahmed {\em et al} have presented a family of solvable potentials that seem to support the conjecture. Our results may be useful for an eventual discussion of that still unproved conjecture.

Another interesting issue raised in Ref.~\cite{nobl} is what could be called the ``existence of a classically allowed region" in complex potentials. Noble {\em et al} have shown that the eigenfunctions of $\mathcal{PT}$-symmetric anharmonic oscillators are concentrated in a region where the eigenenergy is above the modulus of the complex potential. The same fact can be observed in our case of purely imaginary potentials. For instance, in the case of $V_3^{\mathcal{PT}}=-{\rm i}\,x^3$, the first eigenenergy  $E_1=1.156267072$ is larger than the modulus of the potential for $-1.05\lesssim x \lesssim 1.05$. In Fig.~4, which shows the squared modulus of the normalized eigenfunction, a concentration of the probability in the mentioned classically allowed region is clearly seen. For a quantitative comparison, we report in Table 2 the values of $|\psi_1(x)|^2$ at several points.
\begin{figure}
\resizebox{11cm}{!}{\includegraphics{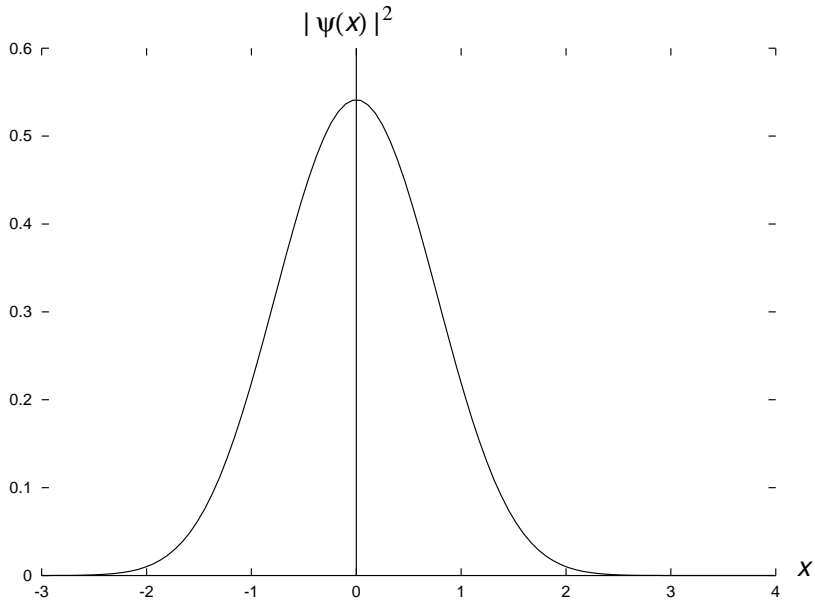}}
\caption{Squared modulus of the normalized first eigenfunction of the Hamiltonian $H_3^{\mathcal{PT}}=-{\rm d}^2/{\rm d}x^2-{\rm i}\,x^3$. The bulk of the probability lies in the region $-1.05<x<1.05$, which would be the classically allowed region, according to Ref.~\cite{nobl}.}
\label{m2wf}
\end{figure}
\begin{table}
\begin{center}
\caption {Numerical values of the squared modulus of the  normalized first eigenfunction of the Hamiltonian $H_3^{\mathcal{PT}}=-{\rm d}^2/{\rm d}x^2-{\rm i}\,x^3$, shown in Fig.~4. The probability is drastically reduced out of the interval (-1.05,1.05), considered by Noble {\em et al} \cite{nobl} as the classically allowed region.}
\begin{tabular}{|l|l|}
\hline
$\pm\,x$ & $|\psi_1(x)|^2$  \\
\hline
0 & 0.541405 \\
0.5 & 0.434904 \\
1 & 0.219137 \\
1.05 & 0.198931 \\
1.1 & 0.179604 \\
1.5 & 0.064408 \\
2 & 0.010063 \\
2.5 & 0.000760 \\
\hline
\end{tabular}
\end{center}
\end{table}

\section{Gamow states}

Once the eigenvalues of the Hamiltonians $H_a^{\mathcal{PT}}$ have been computed, the Gamow energies of the Hamiltonians  $H_a$ are immediately obtained. A table analogous to Table 1, with each entry multiplied by a factor $\exp(-{\rm i}\pi/(a+2))$, would result. Comparison can be made with the values obtained by Fern\'andez and Garcia \cite[Table 4]{fega} for the first six Gamow states of the anharmonic oscillators
\begin{equation}
\frac{1}{2}p^2-x^K\,,  \qquad {\rm with} \quad K = 3 \; {\rm and}\; 5.  \label{vi1}
\end{equation}
Due to the different notation for the kinetic energy term in the Hamiltonians (\ref{i5}) and (\ref{vi1}), our Gamow energies (i. e., the values given in our Table 1 multiplied by $\exp[-{\rm i}\pi/(a+2)]$) for $a=K$ should be multiplied by a factor $2^{-K/(K+2)}$ to be compared with the results given in  \cite{fega}. The agreement is total.

The wave functions of the Gamow states can be written immediately. For small or moderate values of $x$, the expressions in Eq.~(\ref{ii7}), with coefficients related as in Eq.~(\ref{ii8}), are applicable. The fulfilment of Eq.~(\ref{iv5}) requires to take
\begin{equation}
A_1=A\,T_{2,4}^{(+1)}\,,\qquad A_2=-\,A\,T_{1,4}^{(+1)}\,  \label{vi2}
\end{equation}
where $A$ represents an arbitrary constant, which may be chosen conveniently
to fix the norm and the phase of the wave function. For large values of $x$ one should use the asymptotic expressions
\begin{eqnarray}
  \psi_{\rm phys}(x) & \sim & \frac{1}{2}\,A\,\frac{T_{2,4}^{(+1)}}{T_{2,4}^{(-1)}}\,\exp\left[-\frac{(-x)^{1+a/2}}{1+a/2}\right](-x)^{-a/4}\sum_{m=0}^\infty a_{m,3}^{(-1)}\,(-x)^{-m/(2q)}, \nonumber \\  & &  \hspace{220pt} x<0\,,   \label{vi3}  \\
  \psi_{\rm phys}(x) & \sim &  \frac{{\rm i}}{2}\,A\,\exp\left[{\rm i}\,\frac{x^{1+a/2}}{1+a/2}\right]x^{-a/4}\sum_{m=0}^\infty a_{m,3}^{(+1)}\,x^{-m/(2q)},   \qquad x>0\,,   \label{vi4}
\end{eqnarray}
where use has been made of Eqs. (\ref{iii18}) and (\ref{iv4}).

Figures 5 and 6 illustrate the kinds of graphics obtained for the wave functions of the Gamow states of the Hamiltonians $H_a$. We have represented the two lowest eigenstates of $H_6$; the results for other values of $a>2$ are similar. The wave functions are normalized in the form
\begin{equation}
\int_{-\infty}^{+\infty}|\psi(x)|^2\,{\rm d}x = 1\,.  \label{vi5}
\end{equation}
For increasing negative values of $x$, the wave function vanishes according to Eq.~(\ref{vi3}). On the large positive $x$ side, the real and imaginary parts of the wave function oscillate according to Eq.~(\ref{vi4}). The amplitude and the wavelength of the oscillations decrease as $x^{-a/4}$ and $x^{-a/2}$, respectively, as $x$ increases.

It remains to investigate the influence of the existence of Gamow states on the scattering at real energies. With this purpose, we have considered the quantity
\begin{equation}
\Delta\tau=2\,\frac{{\rm d}\,\delta (E)}{{\rm d}E}\,,  \label{vi6}
\end{equation}
proportional to the time delay \cite{caac} suffered by the scattered wave, for the range of energies $-5\leq E \leq 15$. Approximate values of $\Delta\tau$ were calculated by numerical derivation with respect to $E$ of the phase shifts obtained in Section 5. The results for several values of $a>2$ are shown in Fig.~7. Notice that, as said at the end of Section 5, $E$ is not the physical energy, but proportional to it. Consequently, when comparing the different curves shown in the figure, the dependence on $a$ of the horizontal and vertical scales should be borne in mind. In all curves a resonance (peak), associated to the first Gamow state, is clearly marked. It becomes sharper for larger $a$, as was to be expected from the fact that the pole of $S(E)$ in the lower complex half-plane, located at the Gamow energy, approaches the real axis as $a$ increases. For $a\geq 4$, a less marked second resonance, associated to the second Gamow state, becomes apparent. These results suggest that, although the number of Gamow states of $H_a$ seems to be infinite, only a finite number of them have effective influence on the scattering at real energies. The number of true resonances and their mean life increase with $a$.
\begin{figure}
\resizebox{11cm}{!}{\includegraphics{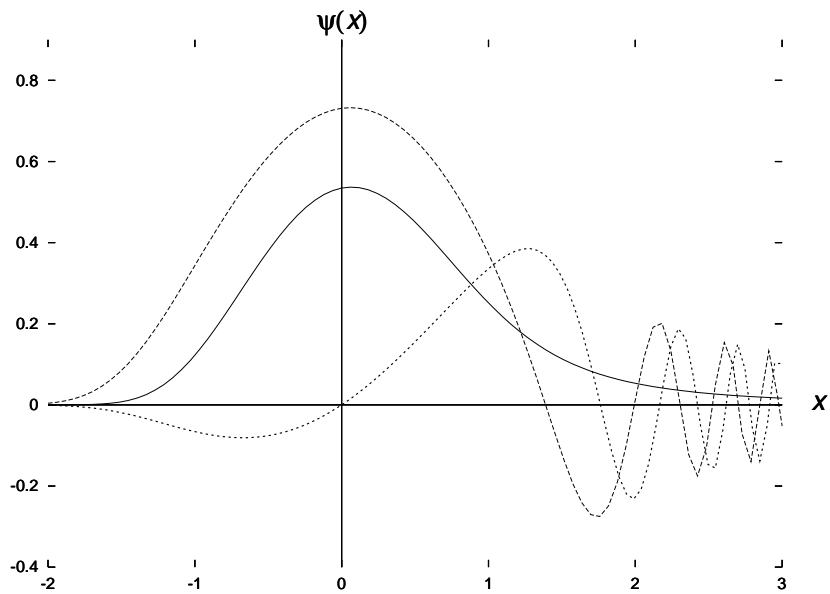}}
\caption{Wave function of the first Gamow state of the Hamiltonian $H_6$. The squared modulus (full) and the real (dashed) and imaginary (dotted) parts of the wave function are shown. The  multiplicative constant has been chosen in such a way that the function becomes normalized as in equation (\ref{vi5}) and positive at $x=0$.}
\label{wfG1}
\end{figure}
\begin{figure}
\resizebox{11cm}{!}{\includegraphics{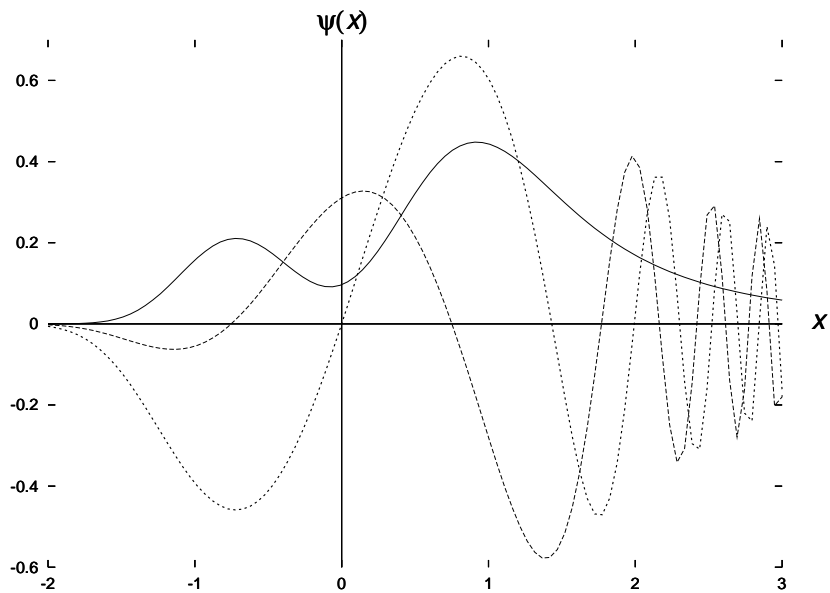}}
\caption{Wave function of the second Gamow state of $H_6$. The comments in the caption of the preceding figure are applicable here.}
\label{wfG2}
\end{figure}
\begin{figure}
\resizebox{11cm}{!}{\includegraphics{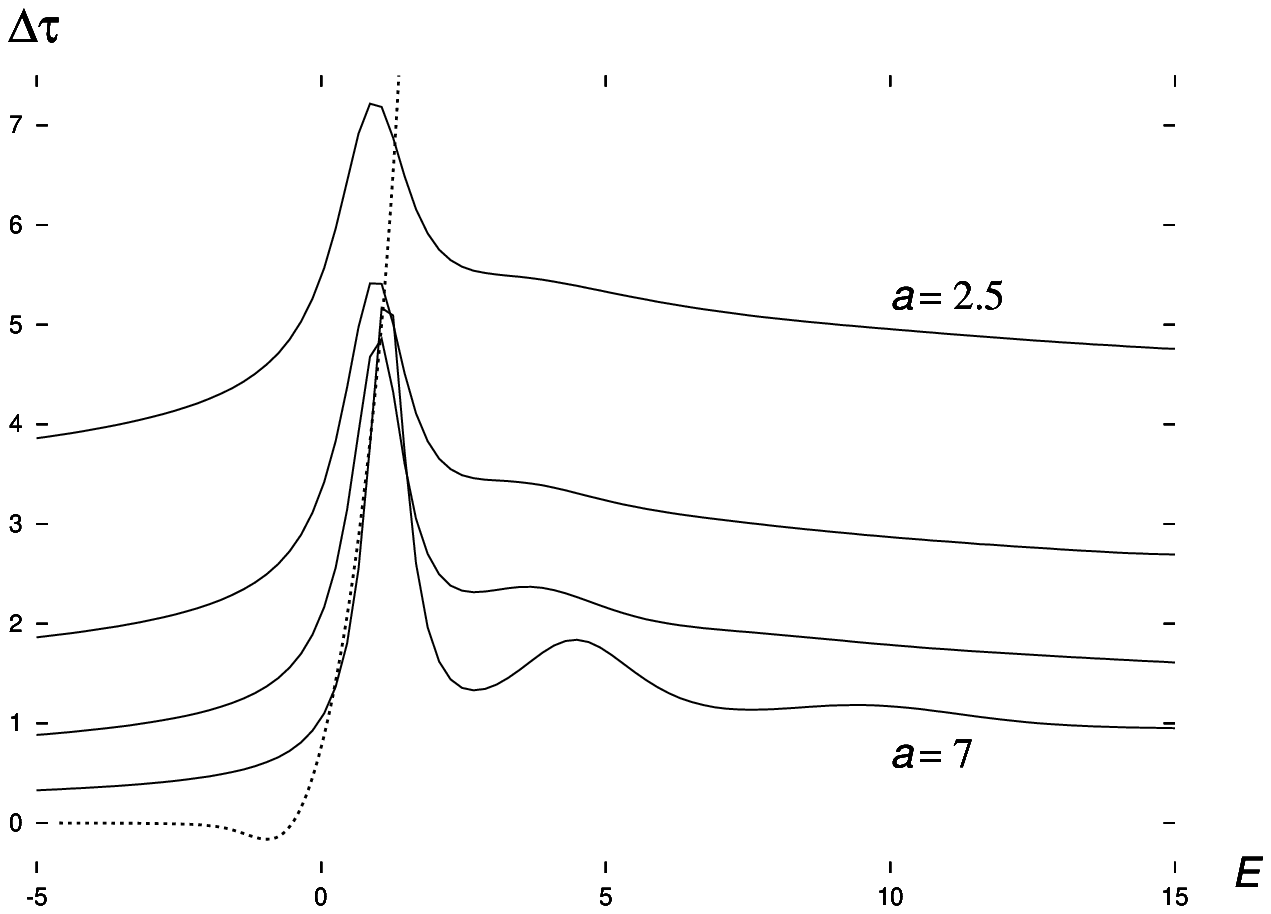}}
\caption{Time delay suffered by a wave scattered by potentials of the form (\ref{i3}) with $a=$2.5, 3, 4, 7 (continuous lines), and 0.5 (dotted line).}
\label{tidea=v}
\end{figure}

\section{Particular case of $a=1/2$}

The formalism developed above, in sections 3 to 7, is directly applicable in the case of $a>2$. Our method of solution of the Schr\"odinger equation, based on the computation of the connection factors of the Frobenius solutions with the Thom\'e ones, is also applicable in the case of $0<a\leq 2$, although general expressions, valid for any $a$, cannot be written. Here we outline the procedure to be followed in the case of $a=1/2$. This is an example that, besides of constituting an illustration of the applicability of our method, gives results very different from those obtained for $a>2$.

The formalism developed in Eqs. (\ref{ii1}) to (\ref{ii8}) is valid also in this case, with $p=1$ and $q=2$. The Thom\'e solutions, however, are not given by Eq.~(\ref{ii9}), but by
\begin{equation}
w_j^{(\sigma)}(t)=\exp \left[\alpha_j^{(\sigma)}\,\frac{t^5}{5}+\beta_j^{(\sigma)}\,\frac{t^3}{3}+\gamma_j^{(\sigma)}\,t\right]\,t^{\mu_j}\,\sum_{m=0}^\infty a_{m,j}^{(\sigma)}\,t^{-m}, \qquad j=3, 4,  \label{vii1}
\end{equation}
with exponents $\alpha_j^{(\sigma)}$ and $\mu_j$ as given in Eqs. (\ref{ii10}) and (\ref{ii12}), new exponents
\begin{equation}
 \beta_j^{(\sigma)}=-\,8E/\alpha_j^{(\sigma)}\,, \qquad  \gamma_j^{(\sigma)}=-\,(\beta_j^{(\sigma)})^2/(2\alpha_j^{(\sigma)})\,, \label{vii2}
\end{equation}
and coefficients $a_{m,j}^{(\sigma)}$ obeying the recurrence relation
\begin{eqnarray}
a_{0,j}^{(\sigma)} =  1, \qquad
2\alpha_j^{(\sigma)}m\,a_{m,j}^{(\sigma)} & = & 2\beta_j^{(\sigma)}\gamma_j^{(\sigma)}\,a_{m-1,j}^{(\sigma)}-2(m\! -\! 1)\beta_j^{(\sigma)}\,a_{m-2,j}^{(\sigma)} \nonumber  \\
 & &\hspace{-150pt} +\,(\gamma_j^{(\sigma)})^2\,a_{m-3,j}^{(\sigma)}-2(m\! -\! 2)\gamma_j^{(\sigma)}\,a_{m-4,j}+\left((m\!-\! 3)(m\! -\! 2)-15/4\right)a_{m-5,j}^{(\sigma)}\,.  \label{vii3}
\end{eqnarray}
Then, Eqs. (\ref{iii1}) to (\ref{iii8}), with $p=1$ and $q=2$, are applicable. Equation (\ref{iii9}), instead, should be replaced by
\begin{eqnarray}
\eta_{n,i,j}^{(\sigma)} &=& \sum_{m=0}^\infty a_{m,j}^{(\sigma)}\Big(\alpha_j^{(\sigma)}\,\hat{c}_{n+m-4,i,j}^{(\sigma)}
+2\beta_j^{(\sigma)}\,\hat{c}_{n+m-2,i,j}^{(\sigma)}+2\gamma_j^{(\sigma)}\,\hat{c}_{n+m,i,j}^{(\sigma)} \nonumber  \\
 & &\hspace{80pt} -\,(n+2m+3+\nu_i)\,\hat{c}_{n+m+1,i,j}^{(\sigma)}\Big),    \label{vii4}
\end{eqnarray}
where the $\hat{c}_{r,i,j}^{(\sigma)}$ are the coefficients of the series expansion of $\exp\left(\beta_j^{(\sigma)}t^3/3+\gamma_j^{(\sigma)}t\right)\,w_i^{(\sigma)}(t)$,
\begin{equation}
\exp\left(\beta_j^{(\sigma)}t^3/3+\gamma_j^{(\sigma)}t\right)\,w_i^{(\sigma)}=\sum_{n=0}^\infty \hat{c}_{n,i,j}^{(\sigma)}\,t^{n+\nu_i}\,.  \label{vii5}
\end{equation}
These coefficients can be calculated by means of the recurrence relation
\[
\hat{c}_{0,i,j}^{(\sigma)}=1\,, \qquad  \hat{c}_{4,1,j}^{(\sigma)} = \gamma_j^{(\sigma)}\left(\beta_j^{(\sigma)}/3+(\gamma_j^{(\sigma)})^3/4!\right)\,,
\]
\begin{eqnarray}
n(n+2\nu_i-1)\hat{c}_{n,i,j}^{(\sigma)} & = & 2(n-1+\nu_i)\gamma_j^{(\sigma)}\,\hat{c}_{n-1,i,j}^{(\sigma)}   -(\gamma_j^{(\sigma)})^2\,\hat{c}_{n-2,i,j}^{(\sigma)} \nonumber  \\
 & & \hspace{-50pt}+\,2(n-2+\nu_i)\beta_j^{(\sigma)}\,\hat{c}_{n-3,i,j}^{(\sigma)}
 -2\beta_j^{(\sigma)}\gamma_j^{(\sigma)}\,\hat{c}_{n-4,i,j}^{(\sigma)}   \nonumber  \\
 & & \hspace{-50pt}-\,(\beta_j^{(\sigma)})^2\,\hat{c}_{n-6,i,j}^{(\sigma)}
 -16E\,\hat{c}_{n-8,i,j}^{(\sigma)}-16\sigma\,\hat{c}_{n-10,i,j}^{(\sigma)}\,.  \label{vii6}
\end{eqnarray}
(Details of how to come at Eqs. (\ref{vii4}) and (\ref{vii6}) can be found in Ref. \cite[Appendix B]{gom2}.) The procedure, then, continues as in the case of $a>2$, Eqs. (\ref{iii10}) to (\ref{iv11}).

The results obtained for the phase shift and the time delay are shown in Figs. 3 and 7, respectively. They are very different from those found for $a>2$. Instead of increasing monotonously with the energy, the phase shift for $a=1/2$ decreases slightly at negative energies $E\lesssim -0.5$ and then increases with a steeper and steeper slope. The corresponding time delay does not show any indication of the existence of resonances.

\section{Final comments}

We have shown, in the preceding sections, how it is possible to write a scattering function (of the complex energy) which represents the interaction of a wave with a potential of the form given in Eq.~(\ref{i3}). This scattering function obeys the usual unitarity condition, which allows the definition, at real energies, of the phase shift. As it occurs in the case of short-range potentials, poles in the lower complex-energy half-plane correspond to Gamow states. The time delay, obtained as the derivative of the phase shift with respect to the energy, has served to analyze the possible existence of real-energy resonant scattering. Our conclusion is that there exist only a few resonances, at energies associated to the real part of the first Gamow poles. The larger the value of the parameter $a$, the more and sharper resonances are found.

Although it was not the main purpose of this paper, we have found a breakdown of the $\mathcal{PT}$-symmetry in the family of Hamiltonians (\ref{i6}) for $a<a_{\rm cr}\leq 3$. Real eigenvalues and the first complex ones are given explicitly in Table 1 for some rational values of $a$ in
the interval $2<a<3$.

Obtainig global solutions of the Schr\"odinger equation is essential for our determination of the scattering function. In principle, two local solutions valid for $x\to +\infty$ or two other for $x\to -\infty$ can be immediately written: they are the Thom\'e solutions, given in Eq.~(\ref{ii9}). Those for $x\to+\infty$ represent, respectively, incoming  and outgoing waves, whereas those for $x\to-\infty$ correspond, respectively, to exponentially decreasing or increasing waves. To comply with the physical conditions, the combination of incoming and outgoing waves, which involves the scattering function $S(E)$, should match with the solution which vanishes as $x\to-\infty$: this determines $S(E)$. But the matching cannot be imposed directly, but trough other local solutions valid for finite $|x|$: the two Frobenius solutions, given in Eq.~(\ref{ii3}). Frobenius and Thom\'e solutions are linked by the connection factors, as shown in Eq.~(\ref{iii1}). The problem of solving the Schr\"odinger equation with the appropriate boundary conditions becomes, in this way, that of computing the connection factors. We have exploited, with this purpose, a procedure which has proved to be useful for the solution of other quantum mechanical problems. The procedure, developed in Section 4, is rigorous and does not require  approximations like perturbation expansions or truncation of the Hilbert space. However, numerical computation is unavoidable to obtain the coefficients of the local solutions, by application of the corresponding recurrence relations, and for summing the series defining the connection factors and the function of the energy whose zeros are the eigenvalues. As a reward, explicit expressions of the wave functions, in the form of convergent series or asymptotic expansions, are obtained.

\section*{Acknowledgments}The comments of two anonymous reviewers have greatly contributed to improve the presentation of this work. Financial support from Conselho Nacional de Desenvolvimento Cient\'{\i}fico e Tecnol\'{o}gico  (CNPq, Brazil) and from  Gobierno de Arag\'on and Fondo Social Europeo (Project E24/1) and Ministerio de Ciencia e Innovaci\'on (Project MTM2012-33575) is gratefully acknowledged.


\begin{thebibliography}{99}

\bibitem{caac} C.A.A. de Carvalho, H.M. Nussenzveig, Phys. Rep. 364 (2002) 83--174.

\bibitem{merz} E. Merzbacher, Quantum Mechanics, Wiley, New York, 1970, pp 110-112, 128.

\bibitem{nuss} H.M. Nussenzveig, Causality and Dispersion Relations, Academic, New York, 1972, pp 110-111.

\bibitem{oroz} O. Rosas-Ortiz, N. Fern\'andez-Garc\'{\i}a, S. Cruz y Cruz. AIP Conf. Proc. 1077 (2008) 31--57, arXiv:0902.4061[quant-ph].

\bibitem{yari} R. Yaris, J. Bendler, R.A. Lovett, C.M. Bender, P.A. Fedders, Phys. Rev. A 18 (1978) 1816--1825.

\bibitem{cali} E. Caliceti, S. Graffi, M. Maioli, Commun. Math. Phys. 75 (1980) 51--66.

\bibitem{alv1} G. Alvarez, Phys. Rev. A 37 (1988) 4079--4083.

\bibitem{jen1} U.D. Jentschura, A. Surzhykov, M. Lubasch, J. Zinn-Justin, J. Phys. A: Math. Theor. 41 (2008) 095302(16pp).

\bibitem{jen2} U.D. Jentschura, J. Zinn-Justin, Appl. Numer. Math. 60 (2010) 1332--1341.

\bibitem{zin1} J. Zinn-Justin, U.D. Jentschura, Ann. Phys. (N.Y.) 313 (2004) 197--267.

\bibitem{zin2} J. Zinn-Justin, U.D. Jentschura, Ann. Phys. (N.Y.) 313 (2004) 269--325.

\bibitem{jen3} U.D. Jentschura, A. Surzhykov, J. Zinn-Justin, Ann. Phys. (N.Y.) 325 (2010) 1135--1172.

\bibitem{prl} U.D. Jentschura, A. Surzhykov, J. Zinn-Justin, Phys. Rev. Lett. 102 (2009) 011601(4pp).

\bibitem{fer2} E.M. Ferreira, J. Sesma, J. Phys. A: Math. Theor. 47 (2014) 415306(21pp).

\bibitem{ahm2} Z. Ahmed, S. Pavaskar, L. Prakash, One dimensional scattering from two-piece rising potentials: a new avenue of resonances, arXiv:1408.0231[quant-ph].

\bibitem{nfer} N. Fern\'andez-Garc\'{\i}a, O. Rosas-Ortiz, SIGMA 7 (2011) 044(7pp).

\bibitem{ahm1}Z. Ahmed, S. Pavaskar, L. Prakash, Eur. J. Phys. 36 (2015) 048001(9pp).

\bibitem{ahm3} Z. Ahmed, L. Prakash, S. Pavaskar, Non-catastrophic resonant states in one dimensional scattering from a rising exponential potential, arXiv:1408.2367[quant-ph].

\bibitem{fega} F.M. Fern\'andez, J. Garc\'{\i}a, J. Phys. A: Math. Theor. 46 (2013) 195301(11pp).

\bibitem{nau1} F. Naundorf, Globale L\"osungen von gew\"ohnlichen linearen Differential\-gleichungen mit zwei stark sigul\"aren Stellen, Doctoral dissertation, University of Heidelberg, 1974.

\bibitem{nau2} F. Naundorf, SIAM J. Math. Anal. 7 (1976) 157--175.

\bibitem{gom1} F.J. G\'omez, J. Sesma, J. Comput. Appl. Math. 207 (2007) 291--300.

\bibitem{gom2} F.J. G\'omez, J. Sesma, J. Phys. A: Math. Theor. 43 (2010) 385302(14pp).

\bibitem{gom3} F.J. G\'omez, J. Sesma, Eur. Phys. J. D 66 (2012) 6(5pp).

\bibitem{sesm} J. Sesma, J. Math. Chem. 51 (2013) 1881--1896.

\bibitem{abra} M. Abramowitz, I. Stegun, Handbook of Mathematical Functions, Dover, New York, 1965.

\bibitem{bart} G. Barton, Ann. Phys. (N.Y.) 166 (1986) 322--363.

\bibitem{fer1} E.M. Ferreira, J. Sesma, J. Phys. A: Math. Theor. 45 (2012) 415302(14pp).

\bibitem{erde} A. Erd\'elyi, Asymptotic Expansions, Dover, New York, 1956.

\bibitem{hard} G.H. Hardy, Divergent Series, Clarendon Press, Oxford, 1956.

\bibitem{ben1} C.M. Bender, S. Boettcher, Phys. Rev. Lett. 80 (1998) 5243--5246.

\bibitem{ben2} C.M. Bender, Contemp. Phys. 46 (2005) 277--292.

\bibitem{ben3} C.M. Bender, Rep. Progr. Phys. 70 (2007) 947--1018.

\bibitem{fer3} E.M. Ferreira, J. Sesma, An algebraically solvable $\mathcal{PT}$-symmetric potential with broken symmetry arXiv:1401.5937[quant-ph].

\bibitem{shin} K.C. Shin, J. Math. Phys. 46 (2005) 082110(17pp).

\bibitem{nobl} J.H. Noble, M. Lubasch, U.D. Jentschura, Eur. Phys. J Plus 128 (2013) 93(13pp).

\bibitem{ben4} C.M. Bender, D.W. Hook, L.R. Mead, J. Phys. A: Math. Theor. 41 (2008) 392005(9pp).

\bibitem{ahm4} Z. Ahmed, D. Ghosh, J.A. Nathan, Phys. Lett. A 379 (2015) 1639--1642.

\end{thebibliography}
\end{document}